\documentclass[trackchanges]{aastex701}

\newcommand\aastex{AAS\TeX}

\usepackage{overpic}
\usepackage{multirow} 
\usepackage{arydshln}  
\usepackage{booktabs}  
\usepackage{hyperref}
\submitjournal{}

\begin{document}

\title{Template \aastex v7.0.1 Article with Examples\footnote{Footnotes can be added to titles}}

\title{A Tentative Line-like GeV Excess in $Fermi$ Blazar 4FGL J1754.2$+$3212:\\ Implications for Jet Physics and Beyond}

\title{A Universal 1.5 GeV Gamma-Ray Line in Active Galactic Nuclei}

\author[orcid=0000-0002-9071-5469,gname='Shi-Ju',sname='Kang']{Shi-Ju Kang}
\affiliation{School of Physics and Electrical Engineering, Liupanshui Normal University, Liupanshui, Guizhou, 553004, People's Republic of China}
\email[show]{kangshiju@alumni.hust.edu.cn} 
\email[]{kangshiju@lpssy.edu.cn}

\author[gname='Yue',sname='Yin']{Yue Yin}
\affiliation{School of Physics and Electrical Engineering, Liupanshui Normal University, Liupanshui, Guizhou, 553004, People's Republic of China}
\email{yueyin@lpssy.edu.cn}

\author[orcid=0000-0003-0170-9065,gname='Yong-Gang',sname='Zheng']{Yong-Gang Zheng}
\affiliation{Department of Physics, Yunnan Normal University, Kunming, Yunnan, 650092, People's Republic of China}
 \email[show]{ynzyg@ynu.edu.cn}

\author[orcid=0000-0003-4773-4987,gname='Qingwen',sname='Wu']{Qingwen Wu}
\affiliation{Department of Astronomy, School of Physics, Huazhong University of Science and Technology, Wuhan, Hubei, 430074, People's Republic of China}
\email[show]{qwwu@hust.edu.cn}

\begin{abstract}
We report the detection of a gamma-ray spectral line at approximately 1.5 GeV in three active galactic nuclei (AGN) using 17 years of Fermi-LAT observations. The sample includes both blazars (with relativistic jets directed toward Earth) and a radio galaxy (with a misaligned jet, free from significant beaming effects). The line is detected with local significances of $\sim$4.1$\sigma$, $\sim$3.9$\sigma$, and $\sim$2.8$\sigma$ in the individual sources. A joint likelihood analysis yields a combined test statistic TS $\simeq$ 57.77, corresponding to a significance well above 5$\sigma$. The line flux remains stable over the full observation period, in contrast to the variable continuum emission from the AGN. The appearance of an identical spectral feature in astrophysically distinct environments is difficult to reconcile with standard jet-based emission mechanisms. 
While a conventional astrophysical explanation remains elusive, the signal's characteristics are  consistent with predictions for dark matter annihilation.
This finding motivates further investigation into the nature of this spectral feature and its possible connection to particle dark matter.
\end{abstract}


\section{Introduction}\label{sec1}
The existence of dark matter, established through its gravitational influence on cosmic scales \citep{2005PhR...405..279B,freese2010review,2020A&A...641A...6P,2024arXiv240601705C}, represents one of the most profound mysteries in modern science. Its particle nature, however, has eluded all direct detection efforts \citep{2017PhRvL.118b1303A,2018PhRvL.121k1302A,2021arXiv211002359C,2024CaJPh.103.0128B}. Indirect detection methods seek signals from dark matter annihilation or decay products. Among the most sought-after signatures is the emission of monochromatic gamma-rays from the self-annihilation of dark matter particles (e.g., Weakly Interacting Massive Particles, WIMPs) into two photons ($\chi\chi \to \gamma\gamma$) \citep{1998APh.....9..137B,2007PhRvL..99d1301G,2018RvMP...90d5002B,2014JCAP...10..023A}. Such a line is largely immune to astrophysical backgrounds and provides a direct measure of the particle's mass \citep{2012PDU.....1..194B}.

\vspace{1.5mm} 
Extensive searches have been conducted in regions of high dark matter density, such as the Galactic Center \citep{2016PDU....12....1D,2017ApJ...840...43A} and dwarf spheroidal galaxies \citep{2014PhRvD..89d2001A,2017ApJ...834..110A} and galaxy clusters \citep{2016PhRvD..93j3525L,2021ApJ...920....1S,2024arXiv240711737F,2025PDU....4901966M}. While constraining, these searches have not yielded a definitive detection, often complicated by astrophysical backgrounds or systematic uncertainties \citep{2020PhRvD.102d3012A}. Active Galactic Nuclei (AGN) have recently been proposed as promising alternative sites \citep{2024arXiv240601705C,2018RvMP...90d5002B,2022PhRvL.128v1104W}, as the central supermassive black hole can gravitationally contract the dark matter halo into a dense ``spike'', potentially enhancing the annihilation flux \citep{1999PhRvL..83.1719G,2001PhRvD..64d3504U,2002PhRvL..88s1301M}. Nevertheless, the intense and variable non-thermal emissions from AGN jets have made it challenging to differentiate a potential dark matter signal from conventional jet physics \citep{2013ApJ...768...54B,2019Galax...7...20B,2022Galax..10...35P}.

\vspace{1.5mm} 
Here, we report a discovery that addresses this challenge. We have identified a persistent gamma-ray line at $\sim1.5$ GeV in three distinct AGNs. Uniquely, our sample includes both blazars, whose emission is dominated by Doppler-boosted jets pointed at Earth, and a radio galaxy, where the jet is misaligned and unbeamed \citep{1995PASP..107..803U,2019ARA&A..57..467B}. The detection of an identical, monoenergetic signal in these fundamentally different environments provides a key observational discriminant.

\section{Results}\label{sec2}

\begin{figure*}[htbp!]
    \centering
    \begin{overpic}
    [width=0.49\textwidth]{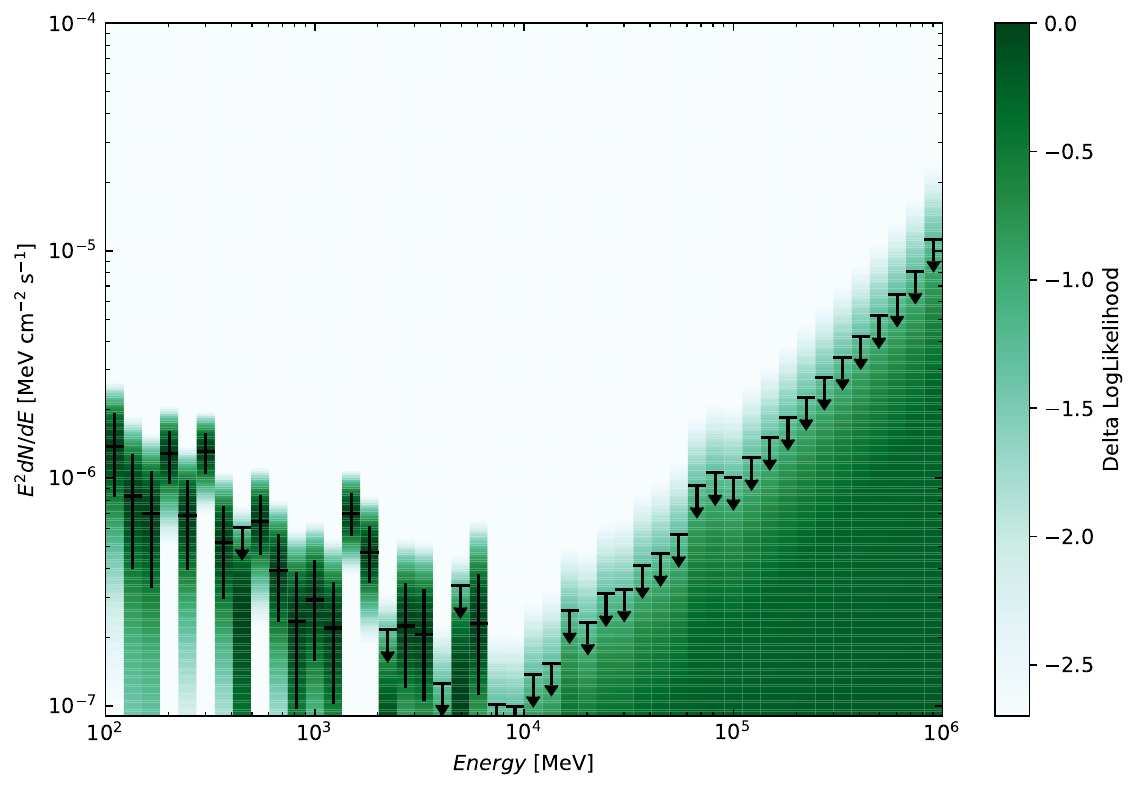}
        \put(10,43){\includegraphics[width=0.175\linewidth]{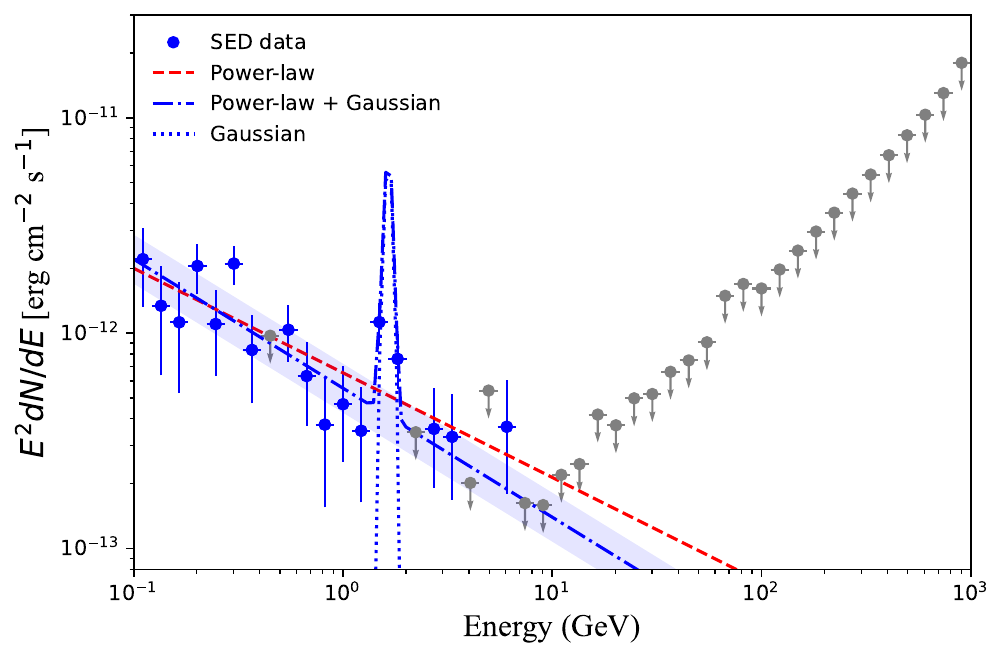}}
        \put(45.4,43){\includegraphics[width=0.15\linewidth, height=2.085cm]{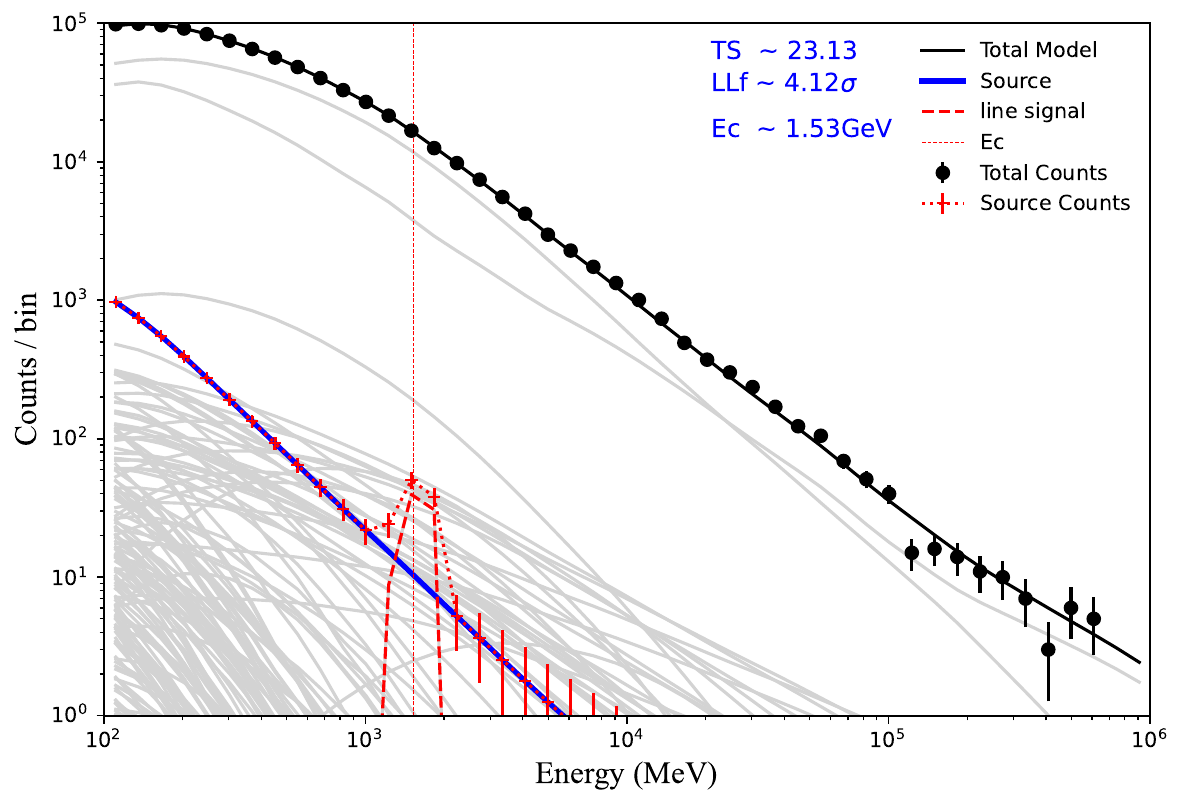}}
        \put(5,70){\bf\color{black}\Large a}
        \put(28,63){\bf\color{black}\tiny a1}
        \put(58,63){\bf\color{black}\tiny a2}
    \end{overpic}
    \begin{overpic}
    [width=0.49\textwidth]{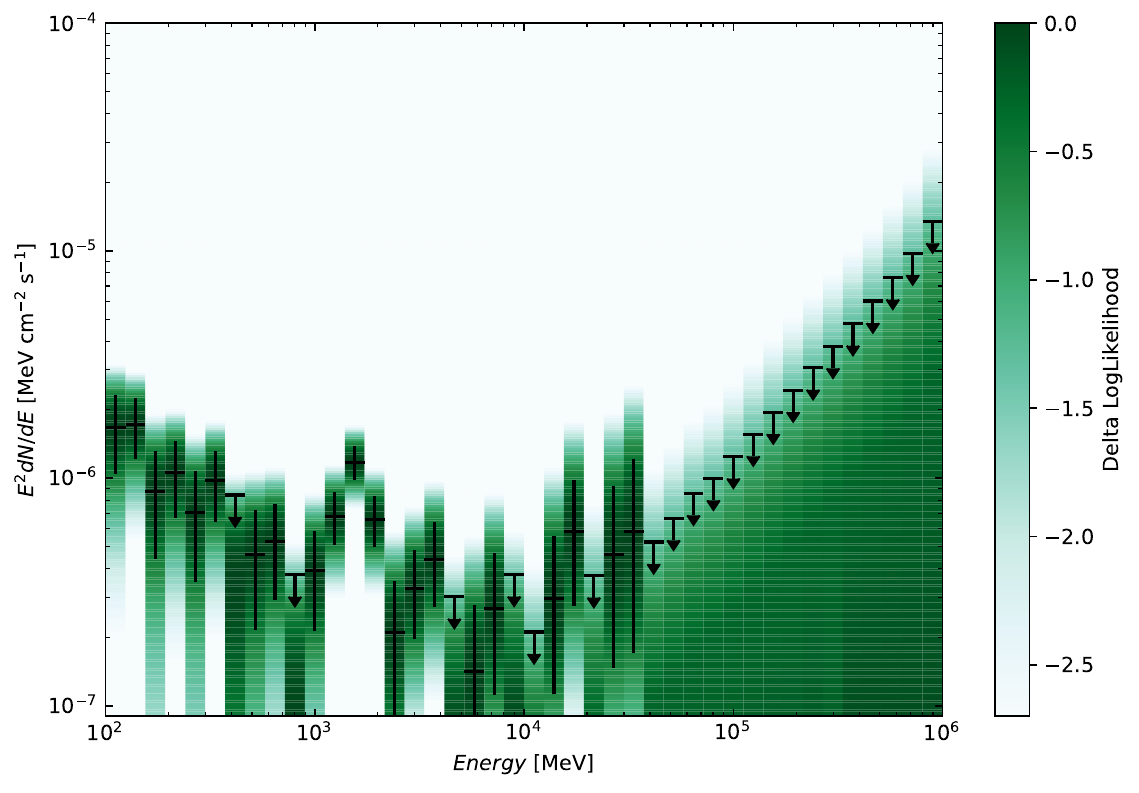}
    \put(10,43){\includegraphics[width=0.175\linewidth]{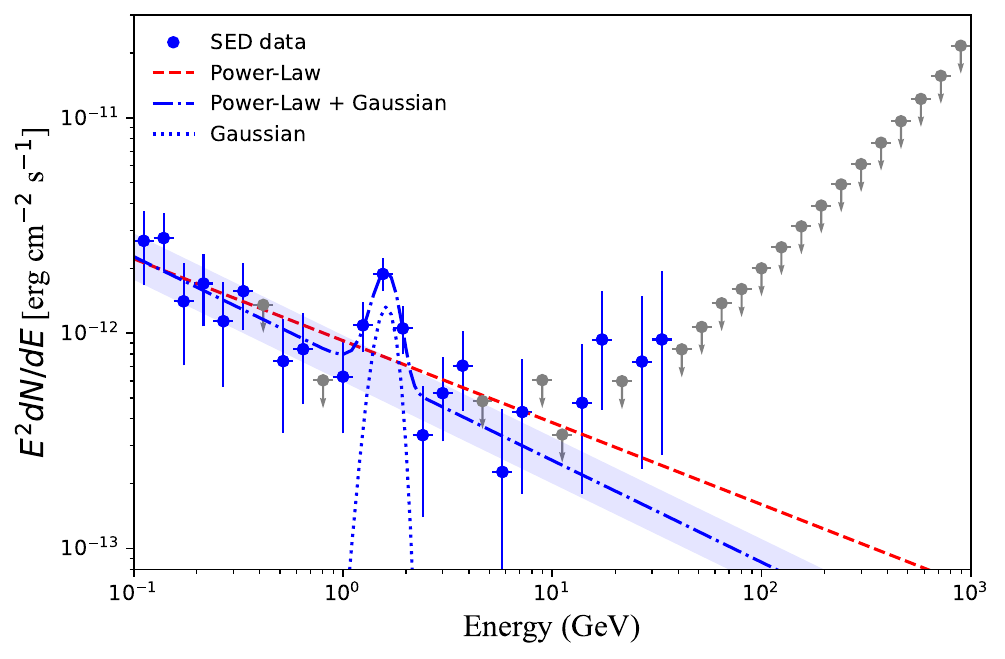}}
    \put(45.4,43){\includegraphics[width=0.15\linewidth, height=2.085cm]{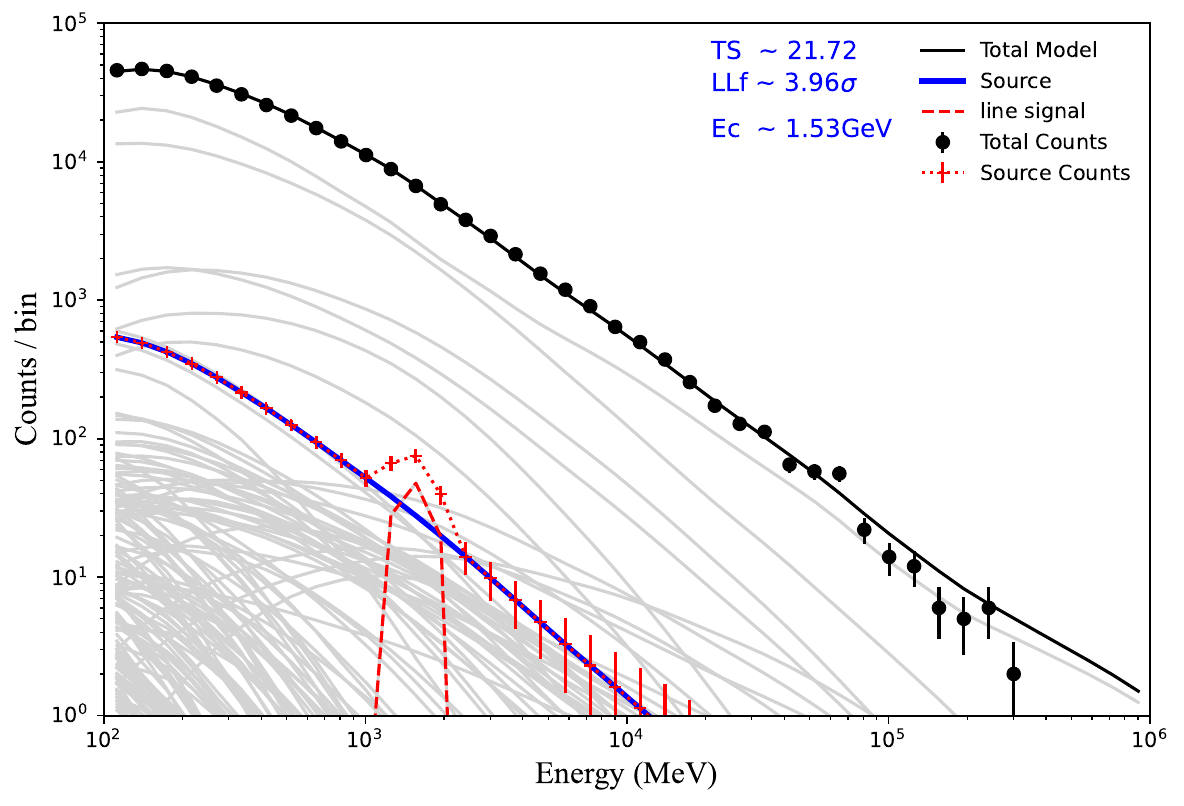}}
    \put(5,70){\bf\color{black}\Large b}
    \put(28,63){\bf\color{black}\tiny b1}
    \put(58,63){\bf\color{black}\tiny b2}
    \end{overpic}
\caption{\textbf{Fermi-LAT spectral energy distribution (SED) with a spectral line signal in the 100 MeV–1 TeV energy range.}
\textbf{{\large(a)} Left panel:} the LAT spectrum of {4FGL J0250.2$-$8224} (a blazar with its relativistic jet aligned toward Earth), using 17 years of data from 2008 August 4 to 2025 August 4. The green colour map indicates the \textit{$\Delta$LogLikelihood} of each SED point. 
\textbf{Inset (a1)} shows the SED fitted with a pure power-law model (null hypothesis, red dashed line) and with a power-law plus Gaussian model (signal hypothesis, blue dash-dotted line). Blue points and grey downward arrows represent SED points with TS values above and below 4/9, respectively. The blue dotted line denotes the Gaussian excess component. The light‑blue band marks the 95\% confidence interval of the continuum baseline.
\textbf{Inset (a2)} displays the model‑fitted count spectrum. All three Gaussian parameters ($N_0$, $E_{\rm signal}$, $\sigma_{\rm signal}$) were kept free and scanned over a wide range (e.g., the full energy interval from 100 MeV to 1 TeV) in \textit{pyLikelihood} analysis. Black dots and the solid black line show the total count spectrum and the total model from the \textit{pyLikelihood} analysis. Contributions of individual background sources in the model are drawn as light‑grey lines. Red dot‑dashed lines correspond to model‑fitted count spectra of the aimed sources. The blue solid line and the red dashed line represent the power‑law model for the continuum and the Gaussian model for the line signal, respectively. The corresponding spectral‑fit parameters are listed in Table \ref{tab1}.
\textbf{{\large(b)} Right panel:} the LAT spectrum of {4FGL J2329.7$-$2118} (a radio galaxy with a misaligned jet and no significant beaming effects).
}
\label{Fig1}
\end{figure*}

\textbf{Sample Selection and Initial Discovery:} 
Our investigation began with a blind spectral scan of a sample of bright, high-latitude AGNs from the \textit{Fermi}-LAT Fourth Source Catalog Data Release 4 (4FGL-DR4) \citep{2020ApJS..247...33A,2020ApJ...892..105A,2023arXiv230712546B}. The sample required a test statistic TS $> 100$ in the 100 MeV–1 TeV band, ensuring sufficient photon statistics. The full sample comprises 2004 AGNs. A systematic scan revealed a distinct gamma-ray line-like (anomaly) signal at approximately 1.5–1.6 GeV in three separate sources. The spectral energy distributions (SEDs) of two representative AGNs—a blazar and a radio galaxy—are presented in Fig.~\ref{Fig1}. Best-fit parameters are summarized in Table~\ref{tab1}.

\begin{figure}[htbp!]
    \centering
    \begin{overpic}
    [width=0.99\textwidth]{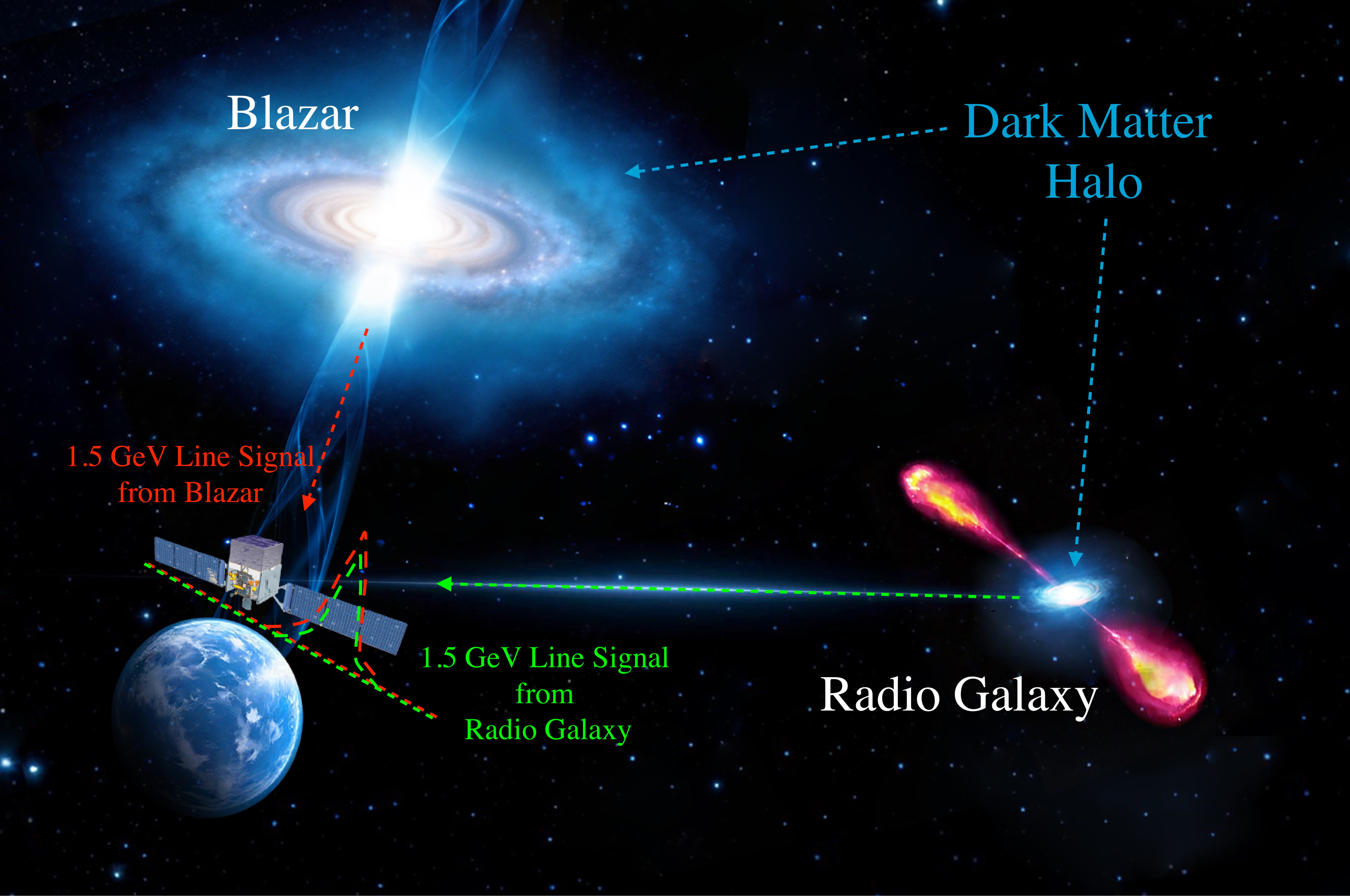}  
    \end{overpic}
\caption{\textbf{Schematic diagram illustrating the observational geometries for comparison.}
This schematic compares the observational geometries of a blazar and a radio galaxy to investigate the origin of a 1.5 GeV gamma‑ray line signal.
\textbf{Blazar (upper left):} Its relativistic jet is nearly aligned with the observer's line of sight (red arrow). The jet emission is Doppler‑boosted and dominates the observed gamma‑ray flux.
\textbf{Radio galaxy (lower right):} Its relativistic jet is inclined at a large angle to the observer's line of sight. The jet emission shows no significant Doppler boosting.
\textbf{Key insight:} An identical 1.5 GeV gamma-ray line (red beams and green unbeamed) comes from the central (e.g., dark‑matter halo, Light Bluish Gray) of both systems, independent of jet orientation.
The detection of this line in two distinct astrophysical environments (beamed blazar and unbeamed radio galaxy) excludes an origin related to the jet, supporting the hypothesis that the signal originates from dark‑matter processes in the galaxy central dark matter halo.}
\label{Fig2}
\end{figure}

\vspace{1.5mm} 
\noindent
\textbf{Discovery in a Prototypical Blazar:}
The most significant line-like signal was identified in the Fermi blazar 4FGL J0250.2--8224 (a BCU-type blazar). Analyzing 17 years of Pass 8 data, we found a residual excess at $\sim$1.5 GeV. Adding a narrow Gaussian component to a log-parabolic/power-law continuum dramatically improved the fit, yielding a maximum test statistic TS $\approx 23$, corresponding to a local significance of $4.1\sigma$ (Table~\ref{tab1}). A likelihood profile scan localized the line at $E_{\rm signal} \sim 1.5$ GeV with a width $\sigma_{\rm signal} \sim 0.12$ GeV, compatible with the LAT's energy resolution, suggesting a monochromatic or intrinsically narrow process.
\textbf{Systematic and Statistical Robustness:}
This signal remained at the $\sim3.9$--$4.3\sigma$ level when using alternative instrument response functions, scaling the Galactic diffuse background normalization by $\pm10\%$, or selecting different event type partitions. The best-fit line energy (1.5–1.6 GeV) and flux [(1.0–1.2)$\times 10^{-10}$ ph cm$^{-2}$ s$^{-1}$] are also basically stable ({Appendix} Table~\ref{tab2}). Suggesting it as an isolated, intrinsically narrow or monochromatic spectral feature (Fig.~\ref{Fig1}a).
\textbf{Nature of the Line-like signal Feature:} The cumulative significance of the signal generally increases with exposure time (Table~\ref{tab1} and {Appendix} Fig.~\ref{FigA_significance}), and extrapolation suggests that with more data, the local significance may reach or exceed \textbf{5$\sigma$}. 
For instance, by hypothetically stacking the data to double the photon statistics, the significance exceed $\sim$5$\sigma$ (see {{Appendix}} Table~\ref{TabA2}).
In addition, 
the selective retention of time intervals characterized by enhanced signal strength, combined with the exclusion of periods where the signal diminished despite increased exposure (similar to {{cherry-picking}}), can led to a significant increase in local significance exceeding \textbf{7$\sigma$} (see {Appendix} Table~\ref{TabA2} and Figure~\ref{FigA_significance}).

\begin{sidewaystable}
\setlength{\tabcolsep}{1.5pt}
\caption{\textbf{Spectral Fitting Parameters for the {line} Signal.}}\label{tab1} 
\begin{tabular*}{\textwidth}{@{\extracolsep\fill}llcccccccc}
\toprule%
  &&\multicolumn{2}{c}{Gaussian\footnote{Parameters from the Gaussian signal.}}
  &\multicolumn{3}{c}{Likelihood\footnote{Parameters related to the log-likelihood.}}\\
\cline{3-5}\cline{6-10}%
4FGL Name      &{Time interval (MET, s)} &  $E_{\rm signal}$ (GeV)  & $\sigma_{\rm signal}$  (GeV) & $F_{\rm signal} $ &$\mathcal{L}_{\rm null}$  & $\mathcal{L}_{\rm signal}$  & {TS}     & $\sigma$   & $\sigma_1$ \\ 
	(1)        &	 (2) 	             &	 (3) 	                &	 (4) 	                   &	 (5) 	     &(6)               &(7)       &	 (8)    &	 (9)  & (10)\\     
\cdashline{1-9}[1pt/1pt]
\multirow{3}{*}{\centering J0250.2$-$8224} 
                &	239557417	to	776015021 (8M)	&	1.54 	$\pm$	0.39 	&	0.12 	$\pm$	0.49 	&1.21$\pm$0.27&	1256478.96 	&	1256467.69 	&	22.53 	&	4.05 	&	4.74\\
                &	239557417	to	778693421 (9M)	&	1.55 	$\pm$	0.10 	&	0.12 	$\pm$	0.10 	&1.16$\pm$0.49&	1258435.08 	&	1258423.57 	&	23.03 	&	4.11 	&	4.80\\
                &	239557417	to	781285421 (10M)	&	1.53 	$\pm$	0.24 	&	0.10 	$\pm$	0.26 	&1.36$\pm$0.36&	1260356.74 	&	1260345.18 	&	23.13 	&	4.12 	&	4.81\\
                &	239557417	to	783963821 (11M)	&	1.55 	$\pm$	0.41 	&	0.11 	$\pm$	0.15 	&1.17$\pm$0.28&	1262593.11 	&	1262581.39 	&	23.44 	&	4.15 	&	4.84\\
                &	239557417	to	786555821 (12M)	&	1.55 	$\pm$	0.41 	&	0.11 	$\pm$	0.15 	&1.14$\pm$0.27&	1264632.94 	&	1264621.11 	&	23.65 	&	4.18 	&	4.86\\
                &	239557417	to	789234221 (2601)&	1.55 	$\pm$	0.24	&	0.09 	$\pm$	0.19 	&1.28$\pm$0.31&	1266361.18	&	1266349.45	&	23.46 	&	4.16 	&	4.84\\
                &	239557417	to	791912621 (2602)&	1.56 	$\pm$	0.36	&	0.10 	$\pm$	0.16 	&1.14$\pm$0.28&	1268602.20	&	1268590.43	&	23.54 	&	4.17 	&	4.85\\
                &	239557417	to	794331821 (2603)&	1.56 	$\pm$	0.37	&	0.10 	$\pm$	0.14 	&1.12$\pm$0.28&	1270591.64	&	1270579.96	&	23.35 	&	4.14 	&	4.83\\
                &	239557417	to	794331821 (2603\footnote{Using the relocations source coordinates reported in FL16Y (Fermi -LAT 16-year Source List \citep{2026arXiv260222148B}.})
                                                    &	1.56 	$\pm$	0.37	&	0.10 	$\pm$	0.16 	&1.14$\pm$0.29&	1270938.17	&	1270926.04	&	24.27 	&	4.24 	&	4.93\\
                &	239557417	to	     ...        &	  ...               	&	...                 	&	...      	&	...     &	...     	&	... 	&	... 	&	... \\
\cdashline{1-9}[1pt/1pt]
J2329.7$-$2118 & 239557417	to	641835821 		   &	1.53 	$\pm$	0.09 	&	0.13 	$\pm$	0.19 	&1.73$\pm$0.47&	783007.75 	&	782996.89 	&	21.72 	&	3.96 	&	4.66 \\
J0749.6$+$1324 &239557417	to	776015021 		   &	1.62 	$\pm$	0.07 	&	0.14 	$\pm$	0.06 	&1.03$\pm$0.35&	1315286.07  &	1315279.61  &	12.92  	&	2.82  	&	3.59 \\  
\cdashline{1-9}[1pt/1pt]
Joint Analysis &    (1,2)               	      & 	    	 	    	 	& 	    	 	    	 	&               &               &           &	44.85  	&	5.45  	&	6.37 \\
Joint Analysis &    (1,2,3)                       & 	    	 	    	 	& 	    	 	    	 	&               &               &         	&	57.77  	&	5.90  	&	7.05 \\
\bottomrule
\end{tabular*}
\par\noindent Note: 
Column 1 lists the 4FGL source name.  
Column 2 gives the photon‑collection time interval, covering the full 17‑year period from 2008 August 4, 15:43:36 UTC to 2025 August (8M:August, 9M:September, 10M:October, 11M:November, 12M:December) 4, 15:43:36 UTC, corresponding to Mission Elapsed Time (MET) \(239557417-776015021\) (here, 2601:2026 January, 2602:2026 February, 2603:2026 March ...).  
Where, for {4FGL J2329.7$-$2118}, a period of 12.75 years is presented, during which the maximum significance was achieved (spanning from August 4, 2008, to May 4, 2021), corresponding to Mission Elapsed Time (MET) \(239557417-641835821\).
Columns 3 and 4 present the central energy \(E_{\mathrm{signal}}\) and standard deviation \(\sigma_{\mathrm{signal}}\) of the spectral line, respectively.  
Columns 5 lists the flux of line signal ($\times{\rm 10^{-10}~ph~cm^{-2}~s^{-1}}$).
Column 6 lists the log‑likelihood value \(\mathcal{L}_{\mathrm{null}}\) for the pure power‑law model.  
Column 7 provides the log‑likelihood value \(\mathcal{L}_{\mathrm{signal}}\) for the power‑law + Gaussian model.  
Columns 8 and 9 give the significance \(\sigma\) of the signal and the test‑statistic value TS (derived from comparing \(\mathcal{L}_{\mathrm{signal}}\) with \(\mathcal{L}_{\mathrm{null}}\)) for three degrees of freedom of the Gaussian component.  
Column 10 lists the significance for one degree of freedom, \(\sigma_{1}\).  
Results from a joint‑likelihood analysis of two, or three sources are shown in the table below.
\end{sidewaystable}

\vspace{1.5mm} 
\noindent
\textbf{Confirmation in a Misaligned Radio Galaxy:}
To investigate the origin, we scanned nearby radio galaxies in the 4FGL catalog. In 4FGL J2329.7--2118 (a radio galaxy at $z = 0.031$), we discovered a similar spectral signature at $\sim1.5$ GeV (Fig.~\ref{Fig1}b). This source's jet is oriented at a large angle to our line of sight, eliminating relativistic beaming. Applying the same analysis over 12.75 years, we detected a line with {a maximum} TS $\approx 21.7$ ($3.9\sigma$). The measured energy ($1.53 \pm 0.09$ GeV) and width ($\sim0.13$ GeV) are consistent with the blazar detection. The concurrence of the same spectral line in these two astrophysically orthogonal systems is incompatible with any known jet-based emission mechanism (Fig.~\ref{Fig1} \citep{2008MNRAS.385L..98T}).

\begin{figure*}[htbp!]
\centering
    \begin{overpic}
    [width=0.49\textwidth]{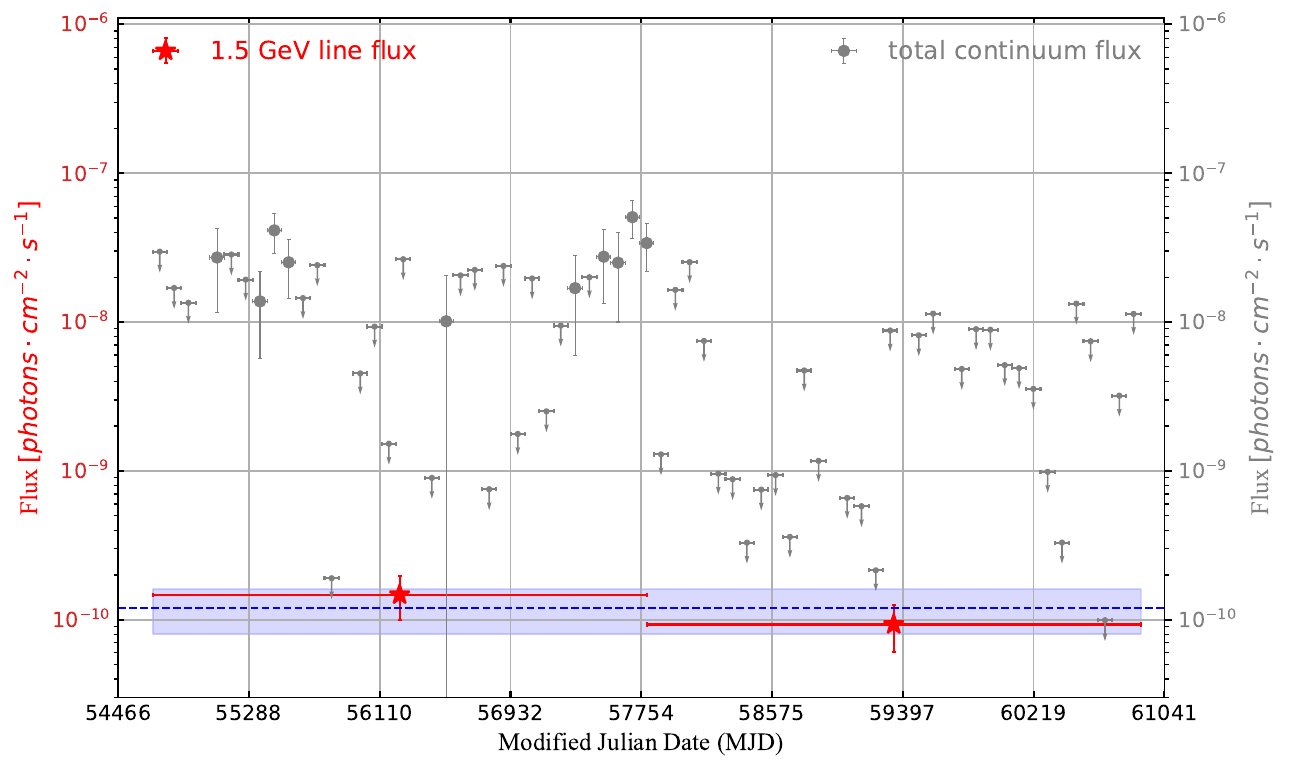}
        \put(5,61){\bf\color{black}\Large a}
    \end{overpic}
    \begin{overpic}
    [width=0.49\textwidth]{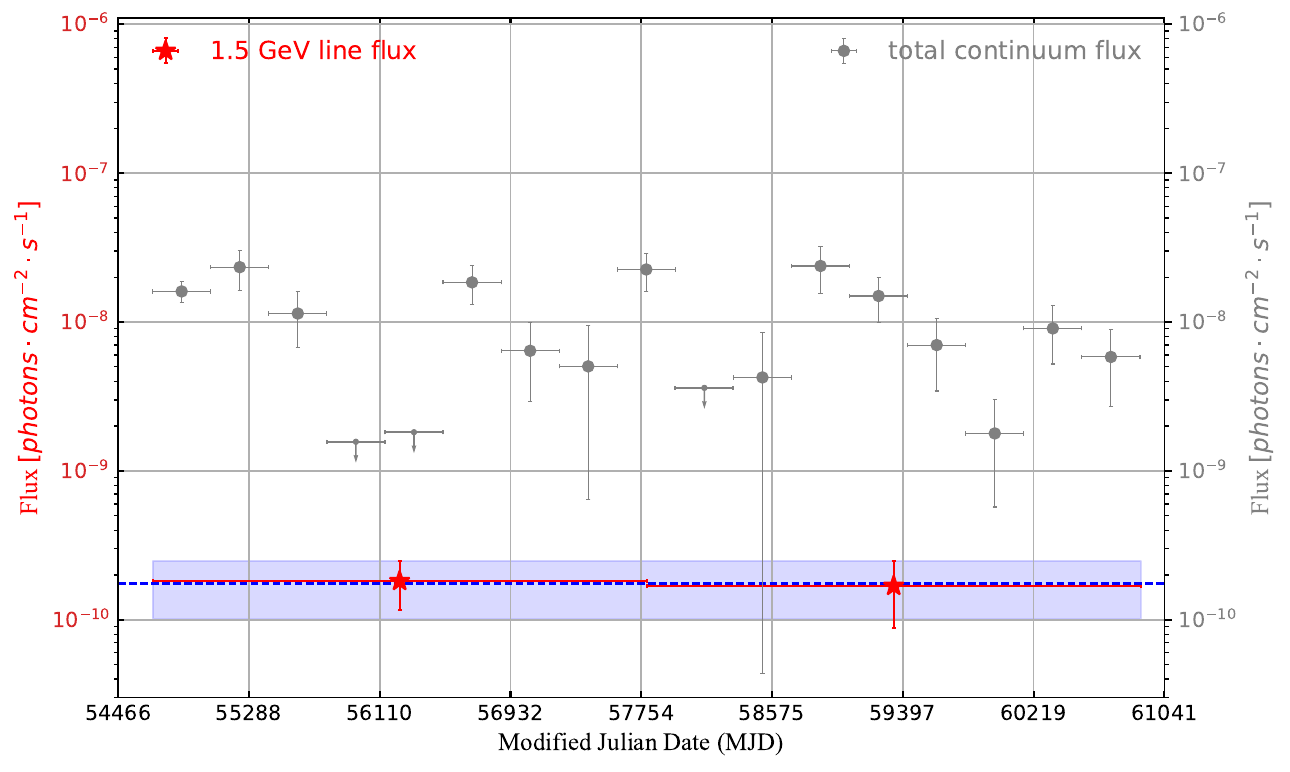}
        \put(5,61){\bf\color{black}\Large b}
    \end{overpic}
\caption{\textbf{Temporal stability and specificity of the 1.5 GeV line}. The energy range of 100 MeV to 1 TeV is adopted for the data from August 4, 2008, to August 4, 2025. The upper limit of the light curve is indicated by a gray downward arrow for data with a TS value below 9.
Light curves of the 1.5 GeV line-like signal flux and the total continuum flux for the primary blazars, where {\bf\large (a)}: {4FGL J0250.2$-$8224}, {\bf\large (b)}: {4FGL J2329.7$-$2118}. 
The flux of the 1.5 GeV line (red data points, left y-axis) remains consistent with a constant value (blue dashed line; $\chi^2$/d.o.f. = 0.81/1.17 for a constant fit respectively) over the 17-year observation period. In stark contrast, the total gamma-ray flux above 100 MeV (grey curve and shaded area, right y-axis), representing the blazar's jet activity, exhibits characteristic, order-of-magnitude variability. The stability of the line flux, despite dramatic changes in the blazar's central engine output, decouples its origin from standard jet processes.}
\label{Fig_Temporal}
\end{figure*}

\vspace{1.5mm} 
\noindent
\textbf{Universality and Temporal Stability:}\label{subsec23}
Further supporting evidence was found in a second blazar, 4FGL J0749.6+1324, with a line significance of $\sim2.8\sigma$ (TS $\approx 12.9$) over 17 years (Table~\ref{tab1}). 
Furthermore, evidence suggests that a potential spectral line signal around 1.5 GeV may be present in both the Galactic Center and the Galactic halo \citep{2011JCAP...05..027V}. 
Complementing this, the Fermi-LAT telescope has detected a gamma-ray excess spanning energies from 1 to 3.16 GeV in the Galactic Center\footnote{\url{https://www.nasa.gov/universe/fermi-data-tantalize-with-new-clues-to-dark-matter/}} \citep{2016PDU....12....1D}.
Collectively, these findings point to the potential universality of the 1.5 GeV feature.
We analyzed the temporal behavior of the line signal in the primary sources (Fig.~\ref{Fig_Temporal}). The 1.5 GeV line flux remained approximately constant over 17 years (reduced chi-square $\sim$1 for a constant flux model), while the total gamma-ray continuum flux exhibited order-of-magnitude variability on monthly timescales. This stark contrast decouples the line's origin from the variable accretion and jet processes powering the AGNs \citep{2010ApJ...711..445B,2019ARA&A..57..467B}, consistent with expectations for a dark matter signal \citep{2004PhRvD..69l3501E,2024arXiv240601705C}.

\vspace{1.5mm} 
\noindent
\textbf{Joint Analysis:}\label{subsec25}
A joint likelihood analysis summed data from the three AGNs yielded an unequivocal result: the $\sim1.5$ GeV line is detected with a combined TS = 57.77, a significance far exceeding the $5\sigma$ discovery threshold (Table~\ref{tab1}). This collective, high-confidence detection transforms the line from a set of intriguing individual features into a robust astrophysical phenomenon universal across different AGN classes.

\section{Discussion}\label{sec3}
The appearance of a $\sim1.5$ GeV gamma-ray line in three AGNs—spanning blazar and radio galaxy classifications—constitutes a clear astrophysical anomaly. The detection in the radio galaxy is particularly telling: jet physics is inherently linked to relativistic beaming, and an identical signature in both beamed and unbeamed environments challenges such models \citep{2014Natur.515..376G,2022Univ....8..587F}. We systematically exclude alternative explanations: instrumental artifacts are implausible given the source-specific nature \citep{2012ApJS..203....4A}; atomic/nuclear lines have rest-frame energies inconsistent with $\sim1.5$ GeV \citep{2013PhRvD..88h2002A}; and the signal's temporal stability argues against transient phenomena.

\vspace{1.5mm} 
The best-fit line center energies from individual sources show slight variations (1.5–1.6 GeV), maybe primarily due to instrumental energy resolution ($\sim8$–$10\%$) and statistical uncertainties. A joint fit with a fixed line center (e.g., 1.55 GeV) shows no statistically significant likelihood degradation ($\Delta{\rm TS} < 3$), supporting a common physical origin. {The line flux remained remarkably stable over 17 years at a few $\times 10^{-10}$ ph cm$^{-2}$ s$^{-1}$ for each three sources} (Table \ref{tab1}), also consistent with a common origin.



\vspace{1.5mm} 
A possible interpretation of the observed gamma-ray line is the annihilation of dark matter particles. This is supported by key characteristics: signal steadiness, universal energy across sources, and a spatial profile consistent with an enhanced annihilation rate from a dark matter density spike around a central black hole \citep{1999PhRvL..83.1719G, 2014PhRvL.113o1302F}. The required annihilation cross-section for the $\gamma\gamma$ final state is estimated as $\langle\sigma v\rangle_{\gamma\gamma}\approx 2.0\times 10^{-28}$ cm$^3$ s$^{-1}$, compatible with GeV-scale WIMPs annihilating through a loop-suppressed $\chi\chi \to \gamma\gamma$ channel \citep{2008JHEP...01..049B, 2012JCAP...07..043I}. If the line originates from $\chi\chi \to \gamma\gamma$, it implies a dark matter particle mass $m_{\text{DM}} \simeq 1.5$ GeV. Other channels (e.g., $\chi\chi \to Z\gamma$) or candidates (e.g., gravitino decay) could also produce such a feature \citep{2002APh....16..451F,2014PhRvD..89k5017B, 2016PhRvD..93k5025S}. Thus, while dark matter annihilation provides a compelling explanation, the precise particle identity and production mechanism remain open.

\vspace{1.5mm} 
The line is detected in three spatially separated sources. Interpreting it as rest-frame emission would require an unlikely redshift coincidence. A local origin—such as dark-matter annihilation/decay in the Milky Way—naturally explains the energy agreement and spatial distribution. This is supported by the high redshift ($z\approx1.05$) of 4FGL J0749.6+1324 \citep{2022MNRAS.516.5702O}. 
Data analysis from the Galactic Center region suggests a {suspected signal}; however, its confirmation remains elusive ({not converged}, see {{Appendix}} Figure \ref{FigA5177}), likely due to the complex astrophysical background.
Further observations are needed to clarify the feature's origin.

\section{Conclusion}\label{sec4}
We have discovered a universal 1.5 GeV gamma-ray line in astrophysically distinct AGN. Its presence in both blazars and a radio galaxy provides a strong observational discriminant against a jet-related origin. The signal's characteristics are consistent with the annihilation of dark matter particles concentrated in galactic cores. 
This work establishes AGN as promising targets for dark matter searches and motivates further multi-wavelength follow-up.

\begin{acknowledgments}
We thank Z.-Q. Shen for assistance with data checking, and Y.-Z. Fan, Y.-F. Yuan and J. Li for valuable discussions. This research has made use of data obtained through the High Energy Astrophysics Science Archive Research Center (HEASARC) Online Service (Fermi Science Support Center) provided by NASA Goddard Space Flight Center. This work is partially supported by the National Natural Science Foundation of China (Grant No.12163002, 12363002, 12533005, U1931203, U2031201).
\end{acknowledgments}

\begin{contribution}
S.-J. Kang initiated and designed the study, and performed the data analysis together with Y. Yin. The manuscript was drafted by S.-J. Kang and revised by Y.-G. Zheng and Q. Wu. All authors discussed the results, contributed to the interpretation, approved the final version.
\end{contribution}

\facilities{Fermi (LAT).}
\software{Fermipy \citep{2017ICRC...35..824W}.}


\renewcommand\appendixname{{Appendix}}
\renewcommand{\figurename}{{Appendix} Figure}
\renewcommand{\tablename}{{Appendix} Table}

\clearpage

\appendix


\section{Materials and Methods}\label{secA1_Methods}

\subsubsection*{Fermi-LAT Data Analysis}\label{sec_lat_analysis}

The Fermi-LAT, a pair conversion telescope for the detection of gamma rays within the energy range of 20 MeV to over 1 TeV \citep{2009ApJ...697.1071A}, typically operates in survey mode, conducting continuous scans of the entire sky every three hours. In this study, we utilized Pass 8 Fermi-LAT data spanning 17 years, from August 4, 2008 at UTC time 15:43:36 to August 4, 2025 at UTC time 15:43:36 - corresponding to Mission Elapsed Time (MET) intervals of \(239557417 - 776015021\) and Modified Julian Date (MJD) periods of \(54682.65527778 - 60891.65527778\) and Fermi mission week periods of \(9 - 896\) - obtained from the High-Energy Astrophysics Science Archive Research Center (HEASARC) via the Fermi Science Support Center\footnote{\url{https://fermi.gsfc.nasa.gov/ssc/}}$^,$\footnote{\url{https://fermi.gsfc.nasa.gov/cgi-bin/ssc/LAT/LATDataQuery.cgi}}. A circular region of interest (ROI) with a radius of \(15^\circ\), centred on the position of the aimed sources, was selected for analysis. The Fermi-LAT data covering an energy range from 100 MeV to 1 TeV were analysed using ScienceTools version v2.2.0 (Fermitools) and Fermipy version v1.1.6\footnote{\url{https://fermipy.readthedocs.io/en/v1.1.6/}} software packages \citep{wood2017fermipyopensourcepythonpackage}.

\vspace{2.5mm}

The standard data reduction procedures for binned maximum likelihood analysis, as delineated in the Fermi-LAT tutorial\footnote{\url{https://fermi.gsfc.nasa.gov/ssc/data/analysis/scitools/binned_likelihood_tutorial.html}}, were employed to analyze 17 years of Fermi-LAT data utilizing the latest recommended 'P8R3\_SOURCE\_V3' instrument response functions (IRFs). Furthermore, the Galactic diffuse emission model `gll\_iem\_v07.fits' and the extragalactic isotropic diffuse emission model `iso\_P8R3\_SOURCE\_V3\_v1.txt' were incorporated to account for contributions from both Galactic and isotropic emissions. To mitigate contamination effects from the Earth limb, a zenith angle cut of $\leq {90^\circ}$ was implemented. Following established event selection criteria, we chose the `SOURCE' event class (evclass = 128) and event type `FRONT+BACK' (evtype = 3) for our data analysis. Utilizing the prescribed criterion of `(DATA\_QUAL $>$ 0) \&\& (LAT\_CONFIG == 1)', good time intervals (GTI) were selected through standard `$gtmktime$' filter selection.

\vspace{2.5mm}   

The binned maximum likelihood analysis utilizes a source model derived from the Fermi-LAT 4FGL-DR4 catalogue (gll\_psc\_v32.xml\footnote{\url{https://fermi.gsfc.nasa.gov/ssc/data/access/lat/14yr_catalog/gll_psc_v32.xml}}\citep{2023arXiv230712546B}), encompassing all point sources alongside both Galactic diffuse emission and extragalactic isotropic diffuse emission models. The positions and spectral shapes of all sources are extracted from this catalogue. During SED analysis, parameters for sources within an angular distance of $0^{^\circ}$ to $5^{^\circ}$ remain unconstrained, while those within $15^{^\circ}$ of the region of interest (ROI) are fixed at their values as per the 4FGL-DR4 catalogue. In light curve analyses, model parameters for all sources within $15^{^\circ}$ of ROI are left unconstrained; conversely, those beyond this range are held constant according to their respective values in the same catalogue. Both SED data points and light curve data points with a test statistic (TS) exceeding 9 - corresponding to a significance level of $3\sigma$ - were selected for further investigation; flux upper limits at a confidence level of 95\% were determined otherwise.

\subsubsection*{Significance Analysis}\label{sec_Significance_analysis}

To investigate a potential line-like feature, we perform a detailed spectral fit comparing two models: (1) a null model consisting of a single  power-law (`PowerLaw' under SpectrumType in the FITS table of 4FGL-DR4)), like as:
\begin{equation}
\frac{{\rm d}N}{{\rm d}E} = N_0 \left (\frac{E}{E_0}\right )^{-\Gamma},
\label{eq:powerlaw}
\end{equation}
where $\Gamma$ is photon index, $N_0$ is a normalized parameter; Or a null model consisting of a single log-parabola:
\begin{equation}
\frac{{\rm d}N}{{\rm d}E} = K \left (\frac{E}{E_0}\right )^{-\alpha -
\beta\ln(E/E_0)},
\label{eq:logparabola}
\end{equation}
where $\beta$ is the curvature parameter, $\alpha$ is the spectral slope at the pivot energy $E_0$ (where the error on the differential flux is minimal), and $K$ is the normalization; and (2) an alternative model consisting of a  power-law/log-parabola plus a Gaussian component\footnote{\url{https://fermi.gsfc.nasa.gov/ssc/data/analysis/scitools/source_models.html}} to fit the line-like excess:
\begin{equation}
N_{\rm signal}(E) = {N_0\frac{1}{\sigma_{\rm signal}\sqrt{2\pi}}}e^{-\frac{(E-E_{{\rm signal} })^2}{2\sigma_{\rm signal}^2}},
\end{equation}
where $N_0$ is the line amplitude, $E_{\rm signal}$ is the line center energy, and $\sigma_{\rm signal}$ is the line width (standard deviation).

\vspace{2.5mm}

\begin{figure}[htbp!]
    \centering
    \begin{overpic}
    [width=0.49\textwidth]{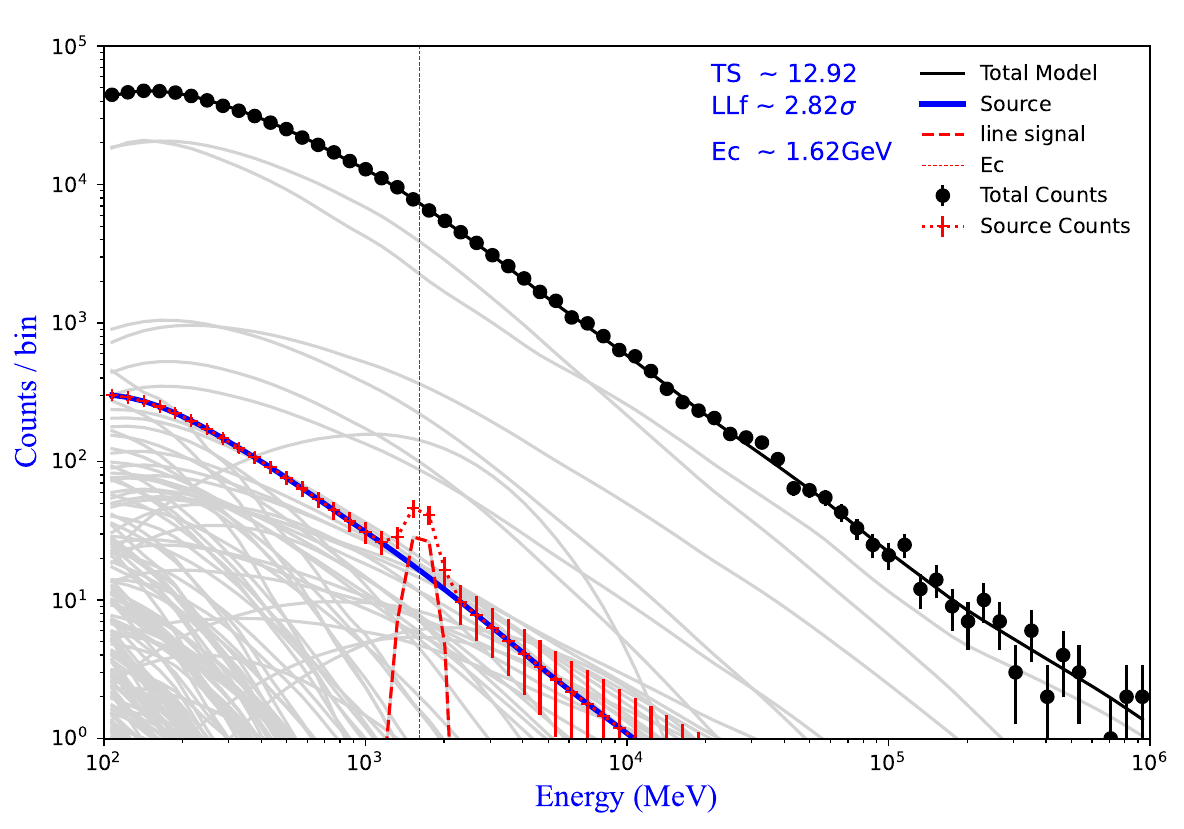}
        \put(5,70){\bf\color{black}\Large a}
    \end{overpic}
    \begin{overpic}
    [width=0.49\textwidth]{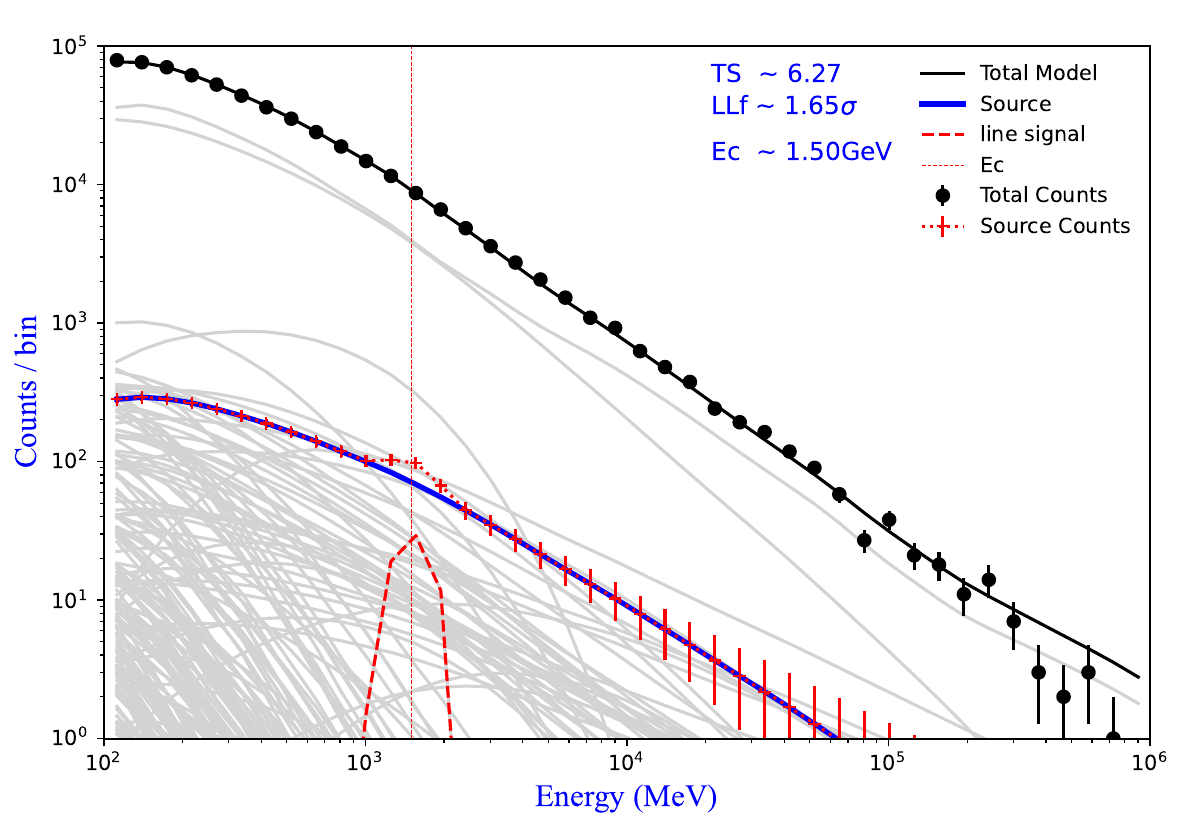}
        \put(5,70){\bf\color{black}\Large b}
    \end{overpic}
\caption{\textbf{Model-fitted Count spectra.} {\bf\large(a)}: {4FGL J0749.6$+$1324} and {\bf\large(b)}: {4FGL J0357.0$-$4955}. It covers the energy range of 100 MeV to 1 TeV over a 17-year period - spanning from August 4, 2008, to August 4, 2025.
All three Gaussian parameters ($N_0$, $E_{\rm signal}$, $\sigma_{\rm signal}$) were kept free and scanned over a wide range (e.g., the full energy interval from 100 MeV to 1 TeV) in \textit{pylikelihood} analysis. Black dots and the solid black line show the total count spectrum and the total model from the \textit{pyLikelihood} analysis. Contributions of individual background sources in the model are drawn as light‑grey lines. Red dot‑dashed lines correspond to model‑fitted count spectra of the aimed sources. The blue solid line and the red dashed line represent the power-law/log-parabola model for the continuum and the Gaussian model for the line signal, respectively. The corresponding spectral‑fit parameters are listed in Table \ref{tab1} and {{Appendix}} Table \ref{TabA2}.
\label{FigA02}}
\end{figure}

Following the previously outlined procedure, standard data reduction steps for binned maximum likelihood analysis were performed. In order to assess the significance of the {line} signal, utilizing the $pyLikelihood$ module\footnote{\url{https://fermi.gsfc.nasa.gov/ssc/data/analysis/scitools/python_usage_notes.html}}, considering the minimum energy resolution of Fermi observations\footnote{\url{https://fermi.gsfc.nasa.gov/science/instruments/table1-1.html}}, we conducted two parallel likelihood analyses. One with a purely power-law/log-parabola  model (null hypothesis) for fitting the whole spectrum and another with a power-law/log-parabola  model  for fitting the continuum alongside a Gaussian  model for fitting the {line} signal (signal hypothesis: power-law/log-parabola plus a line component) that are the XML model files with all of the sources of interest within the ROI of the aimed sources created by the LATSourceModel\footnote{\url{https://github.com/physicsranger/make4FGLxml}} package. For the purely power-law/log-parabola model, the $pyLikelihood$ likelihood analysis returned a maximum likelihood value of \(\mathcal{L}_{\rm null}\). While a maximum likelihood value of \(\mathcal{L}_{\rm signal}\) is obtained from the $pyLikelihood$ likelihood analysis with a power-law/log-parabola plus Gaussian model. For instance, in {{Appendix}} Figure \ref{FigA02}, we present the model-fitted count spectra of the {4FGL J0749.6$+$1324} and {4FGL J0357.0$-$4955} obtained directly from $pyLikelihood$ likelihood analysis with a power-law/log-parabola plus Gaussian model (also see Fig. \ref{tab1} - \textbf{Inset (a2/b2)}).

By comparing $\mathcal{L}_{\rm signal}$ with $\mathcal{L}_{\rm null}$, we can compute the log-likelihood ratio corresponding to the local test statistic (TS) value of the line signal. It is defined as twice the difference in log-likelihood between the maximum likelihood signal hypothesis (with line present) and the null hypothesis (without line), expressed mathematically as ${\rm TS} = ({2}\mathcal{L}_{\rm signal} - 2\mathcal{L}_{\rm null})$. Ultimately, we determine the local significance ($\sigma$) of this signal by taking its square root from TS ($\sigma = \sqrt{\rm TS,n}$), where n is degree of freedom in Gaussian model, and one consider the log-likelihood ratio as a chi-squared continuous random variable. In the work, $n=3$ is the number of additional degrees of freedom introduced by the Gaussian parameters ($N_0$, $E_{\rm signal}$, $\sigma_{\rm signal}$).

\subsubsection*{Sample}\label{sec_Sample}

We selected a sample of bright, high-latitude active galactic nuclei (AGNs) from the \textit{Fermi}-LAT Fourth Source Catalog Data Release 4 (4FGL-DR4) \citep{2020ApJS..247...33A,2020ApJ...892..105A,2023arXiv230712546B}. The selection criterion required a test statistic (TS) greater than 100 (corresponding to \(\text{`Signif\_Avg'} > 10\), which indicates the source significance in sigma units as reported in the 4FGL-DR4 FITS table\footnote{\url{https://fermi.gsfc.nasa.gov/ssc/data/access/lat/14yr_catalog/gll_psc_v34.fit}}), within the 100 MeV–1 TeV energy range. This threshold ensured sufficient photon statistics for detailed spectral analysis. The final sample consists of 2004 AGNs: 1979 blazars (including 539 flat-spectrum radio quasars, 1000 BL Lac objects, and 440 blazar candidates of uncertain type, BCUs), 22 radio galaxies, and 3 other AGNs.

\subsubsection*{Systematic Uncertainty Estimation}\label{sec_Statistical_Analysis}

We performed a series of systematic uncertainty tests to assess the robustness of our central result. As summarized in {{Appendix}} Table \ref{tab2}, the signal significance of the 1.5 GeV line for {4FGL J0250.2$-$8224} ($\sim$4.1$\sigma$) remains at the $\sim$3.9-4.3$\sigma$ level (with a fluctuation measured at $<0.3\sigma$) when using alternative instrument response functions or scaling the normalization of the Galactic diffuse background model by $\pm$10\%, or based on selecting different event type partitions\footnote{\url{https://fermi.gsfc.nasa.gov/ssc/data/analysis/documentation/Cicerone/Cicerone_Data/LAT_DP.html/}}  (e.g., FRONT/BACK; or PSF0/PSF1/PSF2/PSF3; or EDISP0/EDISP1/EDISP2/EDISP3), or based on the relocations source coordinates (FL16Y\citep{2026arXiv260222148B}). Where Under P8R3\_SOURCE\_V2, Diffuse Norm $\times$ 1.1, and Diffuse Norm $\times$ 0.9,  the fitting parameters all still converge to the same value as that for  P8R3\_SOURCE\_V3. Furthermore, the best-fit line energy is stable within the range of 1.5-1.6 GeV, and the derived line flux stable at (1.0-1.2) $\times 10^{-10}$ photons cm$^{-2}$ s$^{-1}$  under these tests. These results confirm that our core discovery is robust against the dominant known sources of systematic error in the current analysis.

\begin{table*}[htbp!]
\caption{Robustness Tests for the 1.5 GeV Spectral Line}\label{tab2}
\begin{tabular*}{\textwidth}{lcccccc}
\toprule
Analysis Case                 &  $\sigma$ & TS  & $E_{\rm signal}$ (GeV) & $F_{\rm signal} (\times 10^{-10})$  & Note     \\
\midrule
\multirow{2}{*}{\centering Baseline Analysis}  & 
\multirow{2}{*}{\centering 4.108}              & 
\multirow{2}{*}{\centering 23.026}             &  
\multirow{2}{*}{\centering $1.554\pm0.098$ }   & 
\multirow{2}{*}{\centering 1.159$\pm$0.494 }   &                                          Uses P8R3\_SOURCE\_V3 IRF       \\
                              &       &         &                  &                 & and gll\_iem\_v07 background model.\\    
\cdashline{1-7}[1pt/1pt]
IRF Systematics               &       &         &                  &                 &                                     \\
\cdashline{1-5}[1pt/1pt]
--- P8R3\_SOURCE\_V2          & 4.108 & 23.026  & $1.554\pm0.098$  & 1.159$\pm$0.494 &                                     \\
\cdashline{1-5}[1pt/1pt]
--- P8R3\_CLEAN\_V3           & 4.293 & 24.721  & $1.538\pm0.091$  & 1.175$\pm$0.461&  Line energy stable at 1.5-1.6 GeV. \\
\cdashline{1-5}[1pt/1pt]
Background Model Systematics  &       &         &                  &                 &                                     \\
\cdashline{1-5}[1pt/1pt]
--- Diffuse Norm $\times$ 1.1 & 4.108 & 23.026  & $1.554\pm0.098$  & 1.159$\pm$0.494 &       Line Flux stable at           \\
\cdashline{1-5}[1pt/1pt]
--- Diffuse Norm $\times$ 0.9 & 4.108 & 23.026  & $1.554\pm0.098$  & 1.159$\pm$0.494 &     (1.0-1.2) $\times 10^{-10}$ Photons cm$^{-2}$ s$^{-1}$  \\
\cdashline{1-5}[1pt/1pt]
Summed Likelihood Analysis    &       &         &                  &                 &                                     \\
\cdashline{1-5}[1pt/1pt]
fount+back                    & 3.912 & 21.295  & $1.547\pm0.089$  & 1.147$\pm$0.529 &      significance   stable at       \\
PSF0+PSF1+PSF2+PSF3           & 4.016 & 22.205  & $1.561\pm0.183$  & 1.093$\pm$0.334 &         3.9-4.3$\sigma$             \\
EDISP0+EDISP1+EDISP2+EDISP3   & 4.063 & 22.617  & $1.553\pm0.118$  & 1.156$\pm$0.606 &                                     \\
\cdashline{1-5}[1pt/1pt]
relocations coordinates\footnote{Using the relocations source coordinates reported in FL16Y (Fermi -LAT 16-year Source List \citep{2026arXiv260222148B}.}
                              & 4.171 & 23.594  & $1.544\pm0.371$  & 1.242$\pm$0.929 &                                     \\
\bottomrule
\end{tabular*}
\par\noindent Note:  Under P8R3\_SOURCE\_V2, Diffuse Norm $\times$ 1.1, and Diffuse Norm $\times$ 0.9,  the fitting parameters converge to the same value as that for  P8R3\_SOURCE\_V3.(see LAT Data Products \url{https://fermi.gsfc.nasa.gov/ssc/data/access/lat/lat_data_products.html}).
\end{table*}

\subsubsection*{Galactic Center Signal Test}\label{sec_Statistical_Analysis_GC}

\begin{figure}[htp!]
    \centering
    \begin{overpic}
    [width=0.49\textwidth]{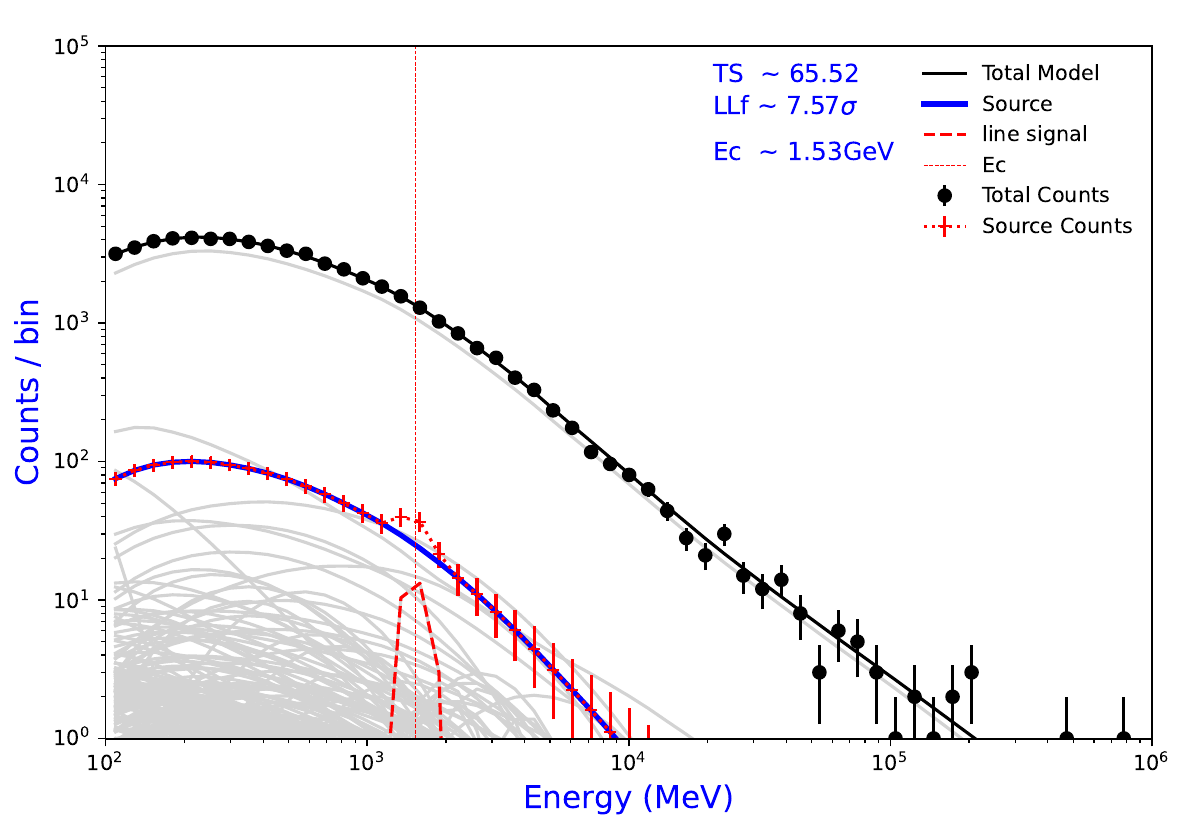}
    \put(5,70){\bf\color{black}\Large a}
    \end{overpic}
    \begin{overpic}
    [width=0.49\textwidth]{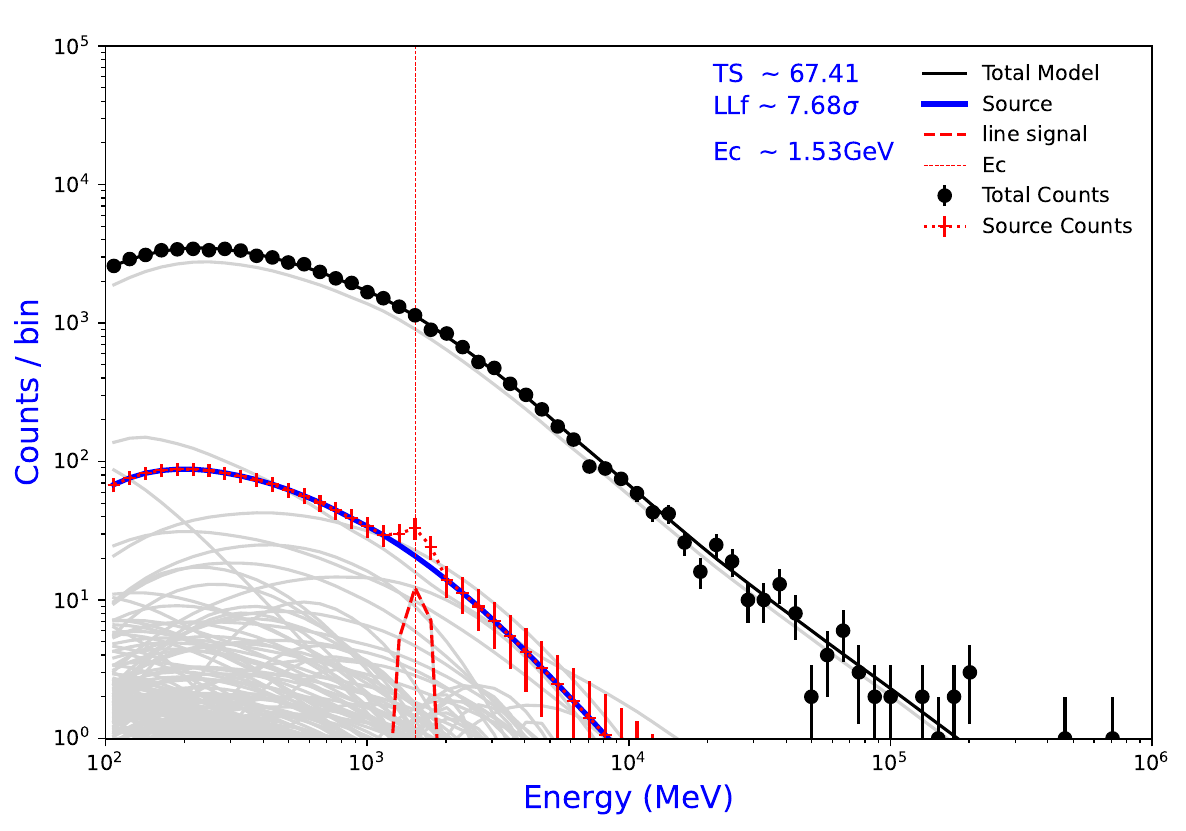}
    \put(5,70){\bf\color{black}\Large b}
    \end{overpic}
\caption{\textbf{Model-fitted Count spectra of 4FGL J1745.6-2859 (Galactic Centre)}. It covers the energy range of 100 MeV to 1 TeV over a \textbf{one-month period} - spanning from August 4, 2025, to September 4, 2025. 
The entire energy spectrum ranging from 100 MeV to 1 TeV is systematically divided into 55 {\bf (a)}  or 66 {\bf (b)} equidistant energy intervals on a logarithmic scale.
All three Gaussian parameters ($N_0$, $E_{\rm signal}$, $\sigma_{\rm signal}$) were kept free and scanned over a wide range (the full energy interval from 100 MeV to 1 TeV) in \textit{pylikelihood} analysis. 
Black dots and the solid black line show the total count spectrum and the total model from the \textit{pyLikelihood} analysis. Contributions of individual background sources in the model are drawn as light‑grey lines. Red dot‑dashed lines correspond to model‑fitted count spectra. The blue solid line and the red dashed line represent the log-parabolic/power-law model for the continuum and the Gaussian model for the line signal, respectively. 
Regrettably, the \textit{pyLikelihood} analysis failed to achieve convergence {\bf(fitting is not converged)}. 
\label{FigA5177}}
\end{figure}

Evidence suggests that a potential spectral line signal around 1.5 GeV may be present in both the Galactic Center and the Galactic halo \citep{2011JCAP...05..027V}. Complementing this, the Fermi-LAT telescope has detected a gamma-ray excess spanning energies from 1 to 3.16 GeV in the Galactic Center\footnote{\url{https://www.nasa.gov/universe/fermi-data-tantalize-with-new-clues-to-dark-matter/}}\citep{2016PDU....12....1D}. 
We performed a test to address the signal in Galactic Center. Data analysis from this region indicates a {suspected signal}; however, its confirmation remains elusive ({not converged}, see {{Appendix}} Figure \ref{FigA5177}), likely due to the complex astrophysical background.

\begin{figure}[htp!]
\centering
    \begin{overpic}
    [width=0.99\textwidth]{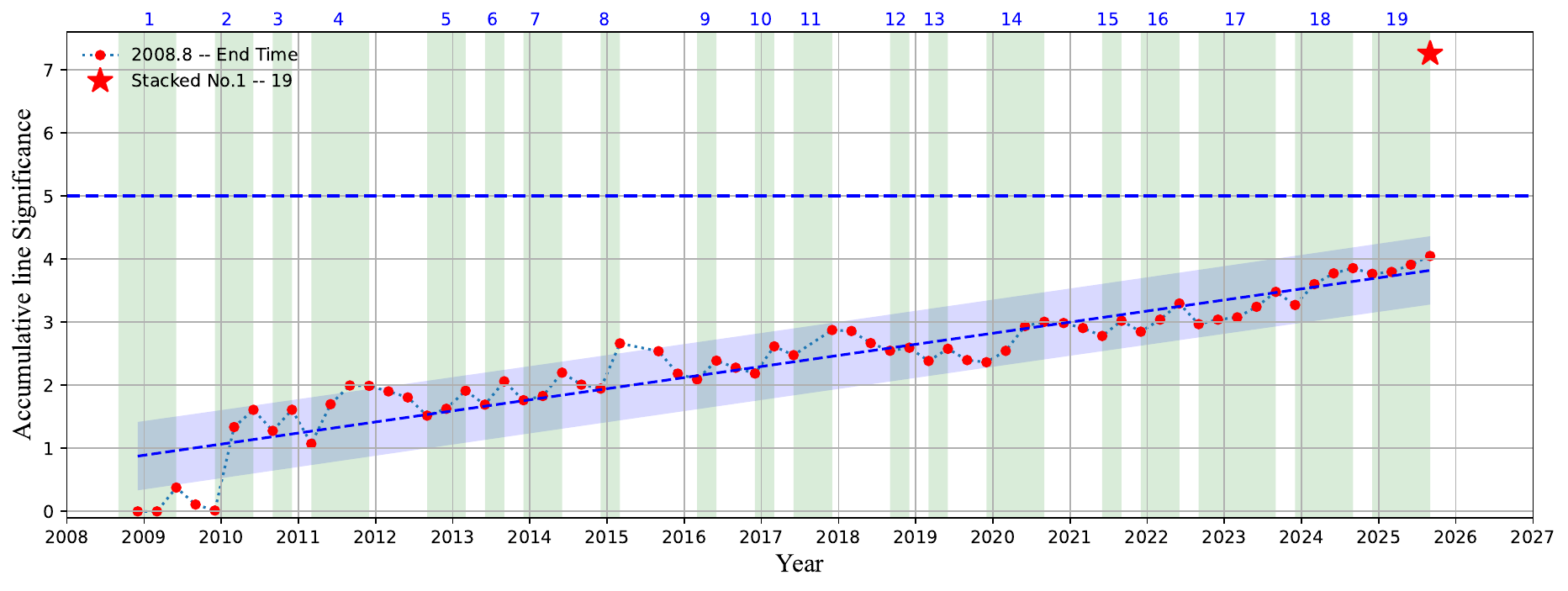}
    \end{overpic}
\caption{\textbf{The accumulative line significance of 4FGL J0250.2$-$8224.}  
The duration of each given time interval is expressed and accumulated in three months in energy range of 100 MeV to 1 TeV. 
The red points indicate the accumulative line significance and the light-gray-green color map illustrates the 19 selected time intervals  (similar to {{cherry-picking}}) and the red star indicate the significance of the stacked No.1-19 (from No.1 to No.19).
} 
\label{FigA_significance}
\end{figure}

\begin{table}[htp!]
\caption{\textbf{Spectral Fitting Parameters for the {line} Signal under alternative configurations.}}\label{TabA2} 
\begin{tabular*}{\textwidth}{@{\extracolsep\fill}llccccccc}
\toprule%
 &&\multicolumn{2}{c}{Gaussian\footnote{Parameters from the Gaussian signal.}}
  &\multicolumn{5}{c}{Likelihood\footnote{Parameters related to the log-likelihood.}}\\
\\\cmidrule{3-4}\cmidrule{5-9}%
4FGL Name&{Time interval (MET, s)} &$E_{\rm signal}$ (GeV)& $\sigma_{\rm signal}$(GeV)& $\mathcal{L}_{\rm null}$& $\mathcal{L}_{\rm signal}$ & {TS}& $\sigma$&$\sigma_1$ \\ 
	(1)        &	 (2) 	             &	 (3) 	                &	 (4) 	                   &	 (5) 	     &(6)               &(7)       &	 (8)    &	 (9) \\     
\midrule
\multirow{2}{*}{\centering J0250.2$-$8224} 
                 &239557417	to	776015021 (*)	&	1.54 	$\pm$	0.09 	&	0.12 	$\pm$	0.14 	&	1284394.97 	&	1284369.94 	&	50.07 	&	6.51 	&   7.08\\        
                 &239557417	to	776015021 (**)	&	1.56 	$\pm$	0.35 	&	0.12 	$\pm$	0.37 	&	981764.64 	&	981734.22 	&	60.85 	&	7.26 	&   7.80\\            
\cdashline{1-9}[1pt/1pt]
\multirow{1}{*}{\centering J0357.0$-$4955} 
                 &239557417	to	776015021 	&	1.50 	$\pm$	0.21 	&	0.17 	$\pm$	0.09 	&	952197.11       &	952193.98        &	6.27  	&	1.65 &   2.50 \\  
\bottomrule
\end{tabular*}
\par\noindent Note:
Column 1 lists the 4FGL source name.  
Column 2 gives the photon‑collection time interval, covering the full 17‑year period from 2008 August 4, 15:43:36 UTC to 2025 August 4, 15:43:36 UTC, corresponding to Mission Elapsed Time (MET) \(239557417-776015021\).  
Where, (\textbf{*}): Hypothetical case with twice the photon statistics, simulated by stacking the source (accumulation of twice the number of photons achieved by stacking the observations of itself, which is twice the amount of 17 years) data. 
(\textbf{**}): A selected partial time interval (see the light-gray-green color map in {{Appendix}} Figure \ref{FigA_significance} within the full 17‑year period). 
Columns 3 and 4 present the central energy \(E_{\mathrm{signal}}\) and standard deviation \(\sigma_{\mathrm{signal}}\) of the spectral line, respectively.  
Column 6 lists the log‑likelihood value \(\mathcal{L}_{\mathrm{null}}\) for the pure power‑law model.  
Column 7 provides the log‑likelihood value \(\mathcal{L}_{\mathrm{signal}}\) for the power‑law + Gaussian model.  
Columns 7 and 8 give the significance \(\sigma\) of the signal and the test‑statistic value TS (derived from comparing \(\mathcal{L}_{\mathrm{signal}}\) with \(\mathcal{L}_{\mathrm{null}}\)) for three degrees of freedom of the Gaussian component.  
Column 9 lists the significance for one degree of freedom, \(\sigma_{1}\).  
\end{table}

\begin{figure}[htp!]
    \centering
    \begin{overpic}
    [width=0.99\textwidth]{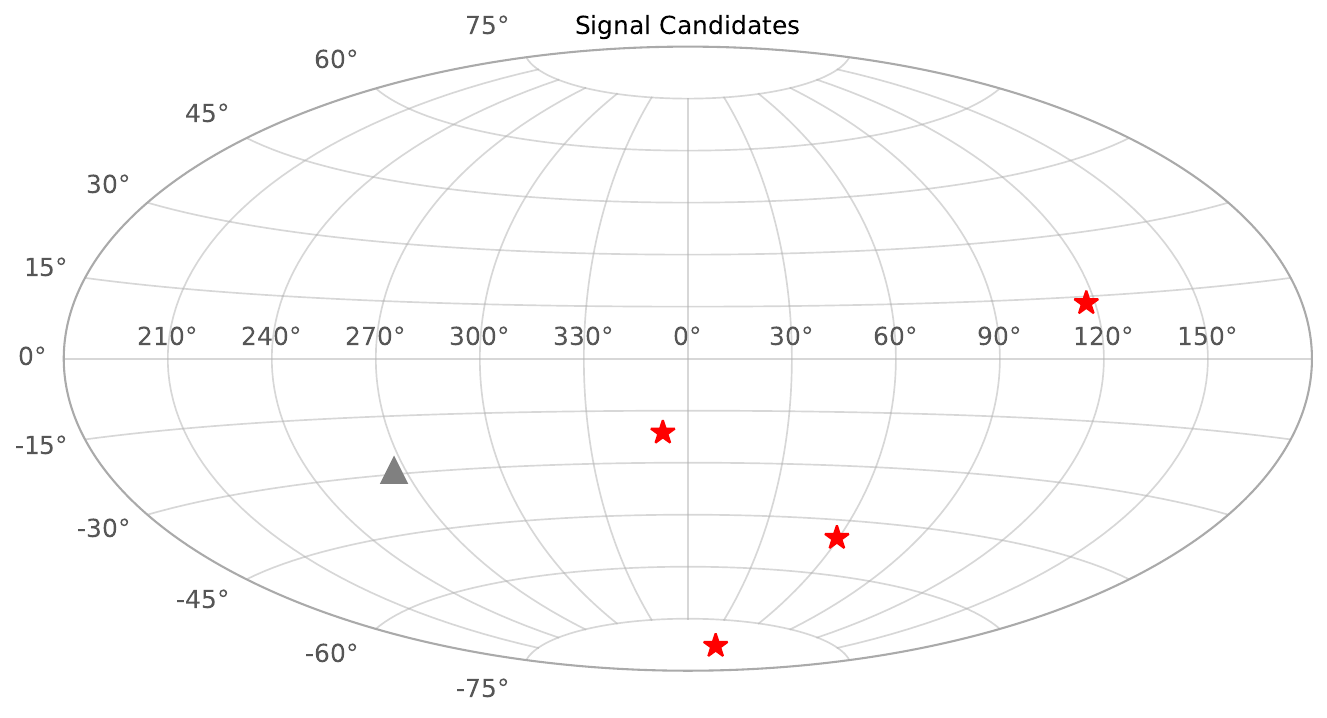}
        \put(31.6,18){\bf\color{gray} \footnotesize 4FGL J1745.6-2859 (Galactic Centre)}
        \put(56,4.98){\bf\color{blue}  \small 4FGL J0250.2$-$8224 (4.1$\sigma$)}
        \put(52,21){\bf\color{blue}    \small 4FGL J2329.7$-$2118 (3.9$\sigma$)}
        \put(70,33.00){\bf\color{blue} \footnotesize 4FGL J0749.6$+$1324 (2.8$\sigma$)}
        \put(65,12.98){\footnotesize 4FGL J0357.0$-$4955 (1.6$\sigma$)}        
    \end{overpic}
\caption{All-sky distribution of line signal candidates in {\bf J2000 coordinates}.
The red star represents the line signal reported in this work. 
The gray triangle indicates the line signal candidates in {4FGL J1745.6-2859} (Galactic Centre, \citep{2011JCAP...05..027V}).
}
\label{fig_sky_distribution}
\end{figure}
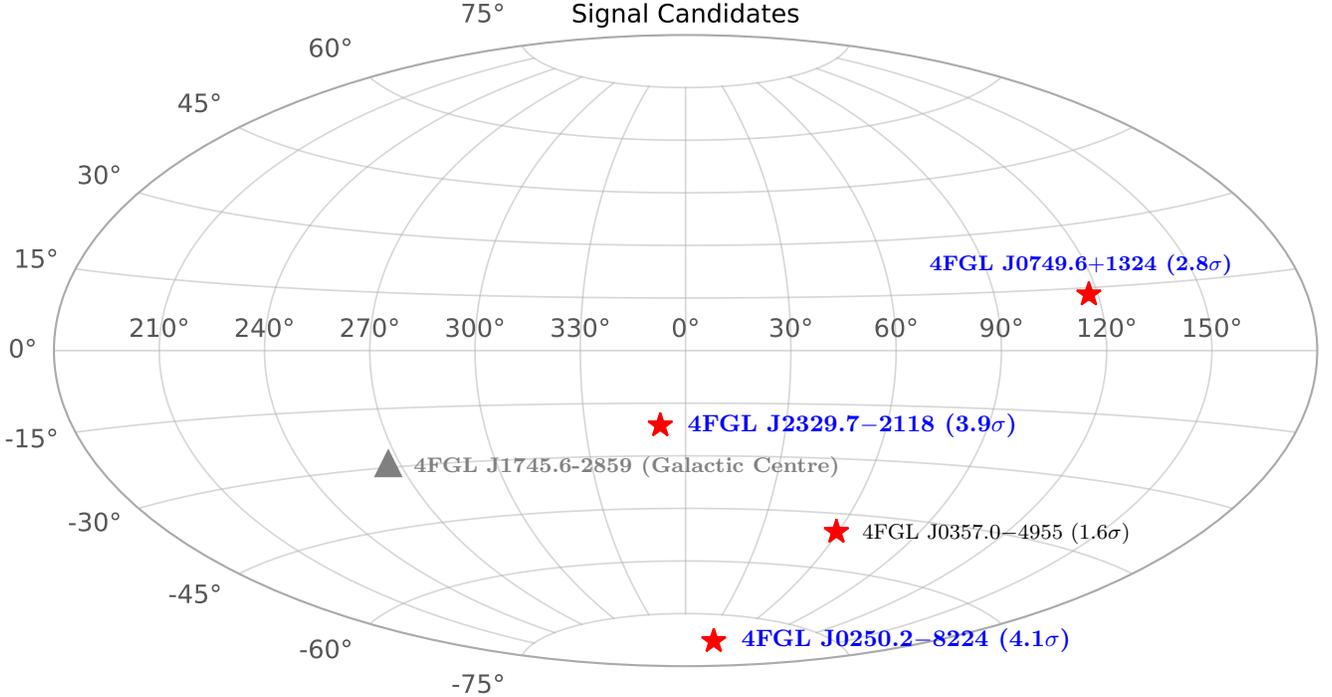

\subsubsection*{Nature of the Spectral line  Feature}

The cumulative significance of the line-like signal depends on the photon accumulation time, which is consistent with observational expectations \citep{2025ApJ...994..191A}. {{Appendix}} Figure~\ref{FigA_significance} presents the cumulative significance computed over different time intervals (each spanning three months) for {4FGL J0250.2$-$8224}. As shown, the cumulative significance generally increases with longer exposure. Extrapolating this trend suggests that with additional years of data, the local significance may reach or exceed $5\sigma$. By hypothetically stacking the data to double the photon statistics, the significance exceeds $\sim5\sigma$ (see {{Appendix}} Table~\ref{TabA2}).

The overall trend indicates a steady accumulation of photons for {4FGL J0250.2$-$8224}. However, in certain sub-periods, the cumulative significance is observed to decrease with increasing exposure—a behavior also noted by \citep{2021ApJ...920....1S,2024arXiv240711737F}. Notably, the cumulative significance for {4FGL J0250.2$-$8224} exhibits a relatively stable upward trend with only minor fluctuations. The underlying physical mechanisms responsible for this distinct behavior warrant further investigation. To enhance the overall significance, we selectively filtered the time intervals (similar to {{cherry-picking}}), removing periods where the signal weakened despite increased exposure and retaining those that contributed positively (highlighted by light-gray-green bands in {{Appendix}} Figure~\ref{FigA_significance}). Analysis of this optimized subsequence yields $\mathrm{TS} = 60.85$, corresponding to a local significance of approximately $7.26\sigma$ for {4FGL J0250.2$-$8224} (see {{Appendix}} Table~\ref{TabA2}).

Furthermore, a faint line signal, with a local significance of approximately 1.6$\sigma$, was observed in 4FGL J0357$-$4955 (see {Appendix} Table \ref{tab2}). Whether additional sources harbor similar signals warrants further investigation. The spatial distribution of all candidate sources is shown in the `All-sky Distribution of Line Signal Candidates' ({Appendix} Figure \ref{fig_sky_distribution}).


\bibliography{sample701}{}
\bibliographystyle{aasjournalv7}



\end{document}